# A Data-Aided Power Transformer Differential Protection without Inrush Blocking Module

Zexuan Lin, Songhao Yang, *Member*, *IEEE*, Yubo Zhang, Zhiguo Hao, *Member*, *IEEE*, Baohui Zhang, *Fellow*, *IEEE*

*Abstract*—When a slightly faulty transformer closes without load, the current waveform presents the coexistence of inrush and fault current. At this time, the inrush blocking module will block the relay, which may delay the removal of the slight fault and lead to more serious faults. To address this problem, this paper proposes a data-aided power transformer differential protection without inrush blocking module. The key to eliminating the negative influence of inrush current is to extract the fundamental component from the non-inrush part of the current waveform, which corresponds to the unsaturation period of the transformer core. Firstly, a data-aided module, namely an Attention module embedded Fully Convolutional Network (A-FCN), is built to distinguish the inrush and non-inrush parts of the current waveform. Then, a physical model of the current waveform is built for the non-inrush part, and the fundamental component is extracted by the nonlinear least square (NLS) algorithm. The proposed method can avoid the block of differential protections when inrush current occurs, which improves the sensitivity and rapidity of the relay, especially in the case of a weak internal fault hidden in inrush current. Finally, simulation and experimental data verify the effectiveness and generalization of the proposed method.

## I. Introduction

POWER transformer connects power generation and distribution, whose state is essential to the safe and stable operation of the power system. Transformer inrush current is a normal phenomenon that occurs when a transformer undergoes a voltage mutation. It's not a fault but the high amplitude may cause the malfunction of transformer protection relays [1]. To this end, the secondary harmonic ratio (SHR) restraint and the dead angle detection methods are widely used in practice to identify the inrush current and block the relay. However, the second harmonic components in inrush current will be decreased with the improvement of core material, and the current transformer (CT) saturation may result in the disappearance of the inrush current's dead angle [2]. Moreover, these detection-blocking strategies leave the transformer in an unprotected state for a short period. Once a fault is hidden in the inrush current, the fault cannot be removed immediately by the relay, which would damage the transformer and even threaten the stability of the system. Therefore, it's essential to propose a transformer differential protection scheme that can handle the negative effects of inrush current.

At present, the inrush current detection methods can be categorized into three kinds: the waveform-feature-based method, the model-based method, and the data-driven method. The first type of method usually uses signal processing technology to extract the time-frequency domain features of current waveforms, e.g., Empirical Fourier transform (EFT) [3], Kalman filtering [4], and Wavelet transform (WT) [5], [6]. These methods can accurately extract certain features in the waveform but are susceptible to noise. Moreover, the thresholds for such methods are usually given empirically and are difficult to rectify. The SHR and dead angle detection also belong to this type. The second type is based on the transformer physical model or the coupling relationship of electrical quantities. Methods in [7] and [8] detect the change of transformer's parameters to judge its state based on the equivalent circuit equation. Literature [9] constructs the equivalent magnetization curve, of which the shape characteristic can reflect the health condition of the transformer. These model-based methods have sufficient theoretical background, but they require high precision of internal parameters and additional measuring devices. Besides, the data-driven methods based on fuzzy logic [10], [11], random forest [12], support vector machine (SVM) [13], artificial neural network (ANN) [14], [15], and probabilistic neural network (PNN) [16], [17] also provide viable solutions to this issue. What's more, some scholars have improved or re-designed the structure of the deep neural network to address this practical problem, making the algorithm faster and more accurate [18], [19]. Reference [20] converts the one-dimensional current into a two-dimensional dynamic differential current image and uses CNN for recognition. These data-driven algorithms have unique advantages in dealing with such nonlinear and multi-factor coupling problems. However, the interpretability and generalizability of artificial intelligence are inherent concerns for the data-driven method.

Since model-based and data-driven methods have their own merits and disadvantages, combining them is an interesting and feasible attempt [21]. Literature [22] utilizes a convolutional neural network (CNN) to identify the unsaturated part of the equivalent magnetization curve proposed in [9]. Literature [23]

This work was partially supported by the Key Research and Development Program of Shaanxi (No. 2022GXLH-01-06), National Natural Science Foundation of China (No. 52007143) and China Postdoctoral Science Foundation (No. 2021M692526). Paper no. TPWRD-00855-2022. (*Corresponding author: Zhiguo Hao*)

The authors are with the Shaanxi Key Laboratory of Smart Grid, Xi'an Jiaotong University, Xi'an 710049, China (e-mail: {405331375, zyb970305}@stu.xjtu.edu.cn; {songhaoyang, zhghao, bhzhang}@xjtu.edu.cn).

proposes a hybrid-driven method in which a data-driven module is used to detect the CT saturation and a physical model is adopted to compensate for the current waveform. These explorations are successful because they tactfully decompose the tasks according to the respective characteristics of the data-driven module and the physical model. Such a deeply combined hybrid-driven idea can also be introduced to address the problem of inrush current detection in transformer protection.

On the other hand, most existing methods adopt the binary classification of inrush current and internal faults. Based on such an assumption, the protection would be blocked once the inrush current is detected. However, internal faults and inrush current can occur simultaneously, e.g., when a faulty transformer is closed without load. If the inrush current covers up the fault current, the protection is blocked and the removal of the internal fault would be delayed.

To avoid the malfunction and false blocking of transformer protection caused by inrush current, this paper proposes a data-aided transformer differential protection without inrush blocking module. The major contributions of this paper are listed below:

1) This paper proposes a data-aided module to detect inrush current. The module is based on an Attention module embedded Fully Convolutional Network (A-FCN), whose function is to identify the inrush and the non-inrush parts of the current waveform. With the aid of the data-driven module, the non-inrush part of the current waveform is retained and the inrush interference is eliminated.

2) The proposed differential protection would not be blocked by inrush current, which is the most significant difference from existing methods. A basic current model and a nonlinear least square (NLS) algorithm are proposed to extract the fundamental frequency component from the non-inrush part of the current waveform. Then the extracted fundamental component is used for differential protection, which prevents protection from failure or false tripping in principle.

This paper is organized as follows. Section II proposes the data-aided inrush identification module. Section III presents the fundamental component extraction module and the overall differential protection scheme. The effectiveness of the proposed protection is verified in section IV by simulation and experimental data. Section V gives the conclusion.

## II. DATA-AIDED INRUSH CURRENT IDENTIFICATION MODULE

### A. Mechanism and Model of Inrush Current

Switching on a transformer without load is the main factor causing inrush current. In the following, the mechanism and model of inrush current are analyzed by taking a single-phase transformer as an example.

In Fig. 1, $r_s$ and $L_s$ are the equivalent resistance and inductance of the power source. $r_{1\sigma}$, $L_{1\sigma}$, $r_m$, $L_m$ represent the resistance and inductance in the primary side and excitation circuit, respectively. Denote $\alpha$ as the switching angle, the following equations are established based on the voltage and flux conservation.

$$\begin{cases} U_m \sin(\omega t + \alpha) = ri_m + \dfrac{d\psi}{dt} \\ \psi = Li_m \end{cases} \tag{1}$$

where $r = r_s + r_{1\sigma} + r_m$, $L = L_s + L_{1\sigma} + L_m$, $i_m$ is the current in the excitation branch, and $\psi$ is the total flux linkage.

Usually, the circuit resistance is much smaller than that of the reactance. Therefore, by ignoring the resistance, the specific expression of $\psi$ is obtained in (2).

$$\psi = -\psi_m \cos(\omega t + \alpha) + \psi_m + \psi_r, \tag{2}$$

where $\psi_m = LU_m / \sqrt{r^2 + (\omega L)^2}$, $\psi_r$ denotes the remanence magnetism.

During normal operation, the transformer's flux $\psi$ is less than saturation magnetic flux $\psi_s$, then $i_m \approx 0$. Once $\psi > \psi_s$, the core saturation leads to rapid decrease of excitation impedance, then inrush current occurs. According to the above analysis, the expression of inrush current is shown in (3).

$$i_m = \begin{cases} \dfrac{U_m}{\omega L}\left[-\cos(\omega t + \alpha) + \cos\alpha + \dfrac{\psi_r - \psi_s}{\psi_m}\right] & \psi \geq \psi_s \\ 0 & \psi < \psi_s \end{cases} \tag{3}$$

For power transformer differential protection, the internal fault current should be distinguished from inrush current. The internal fault current is mainly composed of high-amplitude fundamental component and decaying dc component, whose mathematical model is provided in (4).

$$i_s = A_s \cos(2\pi f_1 t + \alpha) + D_s \exp(-t/T_s), \tag{4}$$

where $A_s$, $D_s$ are the amplitudes of fundamental frequency and dc component, $f_1$=50Hz and $T_s$ denotes the decay time constant.

Furthermore, if a transformer with a slight fault is switched on, the current waveform is presented as the coexistence of fault and inrush current, the magnetization curve and current waveform are shown in Fig. 2.

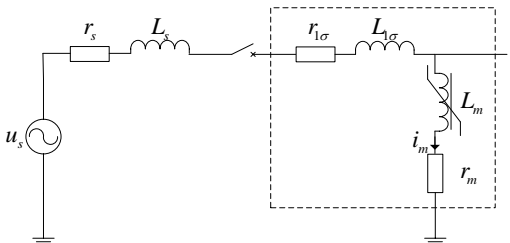

Fig. 1. Equivalent circuit of switching a transformer without load.

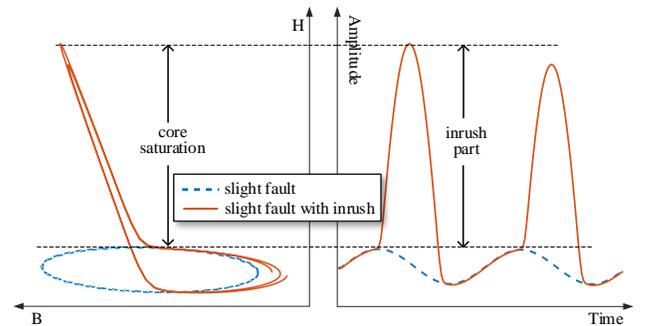

Fig. 2. curve of (a) iron core magnetization and (b) fault current with inrush.

Normally the magnetization curve is a flat ellipse with the long axis perpendicular to H axis [9]. However, as shown in the blue dotted line in Fig. 2, when the transformer exists a weak fault, the short axis is widened and an inclination angle is produced. The current waveform is a typical fault sine wave, and the amplitude is proportional to the magnetic field intensity H. If this faulty transformer is closed without load, the excitation curve is periodically biased to one side when the transformer is saturated, and the current of the corresponding period increases rapidly. The mathematical expression of the current can be obtained by linear addition of (3) and (4). According to (3), the inrush current $i$ of unsaturation part ($\psi \leq \psi_s$) is 0, so the total current is still equal to the actual fault current, which can also be seen from the coincidence relationship between the two curves in Fig. 2(b).

Since the saturation of transformer iron core is periodic, there must exist an unsaturated linear stage in a period. This stage, namely the non-inrush part, has no inrush interference, so it can accurately reflect the true state of the power transformer.

*B. Structure of A-FCN*

According to the above analysis, the inrush part exhibits complex nonlinearity and reduces the accuracy of fundamental component estimation. By contrast, the non-inrush part can reflect the true state of the transformer. Therefore, this paper proposes a data-driven module to eliminate the negative effects of inrush current, whose specific function is to identify the inrush and non-inrush parts of the current waveforms. Then the non-inrush part is retained while the inrush interference is eliminated.

In this paper, FCN is adopted due to its two advantages: 1) high computation efficiency. Each network layer in FCN performs convolution operation sharing filters to form multiple feature maps, which can avoid the repeated calculation of adjacent pixel blocks and improve the computational efficiency. The high computation efficiency of the FCN facilitates its application in real-time protection. 2) flexible adaptability to variable input length. FCN replaces the fully connected layers with convolution layers in the traditional CNN, so it can handle any input current with variable length.

Moreover, Convolutional Block Attention Module, namely CBAM, is introduced to enhance the accuracy and generalization of the FCN. CBAM contains a channel attention module (CAM) and a spatial attention module (SAM). CAM gives high weight to the detailed characteristics from the channel dimension while SAM focuses more on the overview of feature maps from the spatial dimension [24]. The application of CBAM in FCN can help the network focus on key information hidden in massive data.

In general, more FCN layers and more CBAM modules indicate stronger non-linear fitting ability and better accuracy, but at the cost of over-fitting risk and heavy computation burden. Therefore, comprehensively considering the calculation time and performance, this paper constructs the A-FCN with five convolution layers and two CBAMs, as shown in Fig. 3. Noted that the proposed FCN has no pooling layers. The pooling layers are usually adopted in FCN to reduce the size of the network model, but they are not necessary for such a low-complexity network and one-dimension input current. The advantage of no pooling layers is that the detailed relationship between the sampling points can be completely preserved. Besides, the network uses *Adam* optimizer to adaptively adjust the learning rate, and the initial learning rate is set to 0.001 [25]. Meanwhile, L2 regularization ($\lambda_2=0.001$) is adopted to prevent model overfitting, and the activation function of A-FCN is sigmoid. Other parameters of the A-FCN can refer to [26].

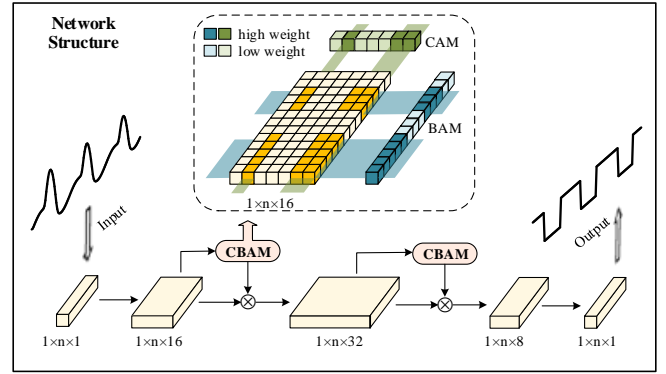

Fig. 3. Specific structure of the proposed A-FCN

*C. Input and Output of the A-FCN*

A-FCN performs end-to-end training without other feature extraction. The input of this network is the measured current $i$ from both sides of the power transformer, which only needs to be normalized as follows.

$$\begin{cases} m = \max(|i|) \\ i_{norm}[k] = i[k]/m \end{cases}, \quad k=1,2,\cdots n, \quad (5)$$

where $m$ denotes the normalization coefficient, i.e., the amplitude of the input current; $i_{norm}$ is the normalized current; $n$ is the total number of sampling points. In the practical application of A-FCN, the time window is one cycle, so the current needs to be normalized per cycle.

Aiming to retain the non-inrush part of the current waveform corresponding to the unsaturation part of the transformer core, the network output is the identification results of the inrush and non-inrush parts. In order to label current waveforms automatically, the labeling method shown in (6) is used.

$$S[k] = \begin{cases} 1, & |i_{norm}[k] - i_{l\_n}[k]| \leq \delta \\ 0, & |i_{norm}[k] - i_{l\_n}[k]| > \delta \end{cases}, \quad (6)$$

where $S$ represents the saturation information sequence of the transformer; $i_{l\_n}$ is the normalized ideal current $i_{l\_n}[k] = i_l[k]/m$, where $i_l$ is the ideal current which can be obtained by an ideal transformer model that would never be saturated. If the transformer core isn't saturated at the $k$-th sampling points, $S[k]=1$; otherwise $S[k]=0$. $\delta$ is the threshold to distinguish the saturated and unsaturated parts in the training cases. Since the label is obtained by the difference between the actual current and ideal current, and the simulation samples contain a large number of saturation or fault situations, the relationship between the magnitudes of the two varies in a wide range. Therefore, this paper proposes a floating threshold to deal with different situations. The threshold $\delta$ can be determined according to Table I, which is based on the amplitude ratio of the actual current to the ideal current. With the floating threshold, the marking error can be minimized as much as possible, and the information about the unsaturation part can be preserved, which balances sensitivity and reliability.

TABLE I
VALUES OF THE FLOATING THRESHOLD

| | $\frac{\max|i|}{\max|i_l|} < 3$ | $3 \leq \frac{\max|i|}{\max|i_l|} < 10$ | $10 \leq \frac{\max|i|}{\max|i_l|} < 20$ | $\frac{\max|i|}{\max|i_l|} > 20$ |
|---|---|---|---|---|
| $\delta$ | 0.03 | 0.01 | 0.005 | 0.003 |

Note that the labeling process is only done when preparing the training and testing datasets for the neural network. In practical applications, the trained A-FCN can automatically mark the unsaturation part of the input current, so it's only necessary to normalize the input current following (5).

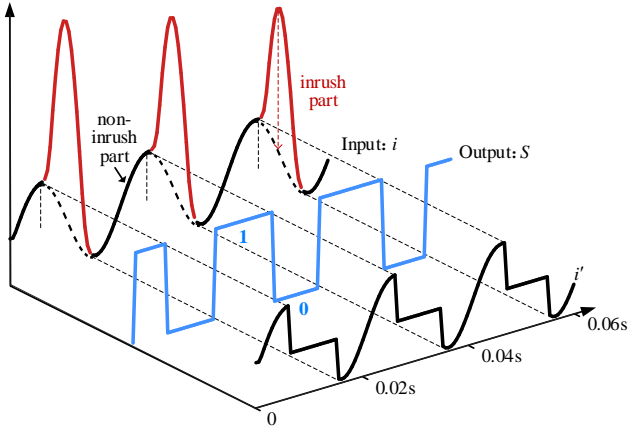

Fig. 4. Input and output of A-FCN.

The input and output information of A-FCN is shown in Fig. 4, by multiplying the output label $S$ and input current $\hat{i}$, the non-inrush part $i'$ of the transformer current can be obtained. $i'$ only retains the sinusoidal fundamental component under unsaturation, so as to eliminate the interference of inrush current.

## III. DIFFERENTIAL PROTECTION WITH HIGH TOLERANCE TO INRUSH CURRENT

### A. Extraction of Fundamental Component

After the process of A-FCN, no matter what kind of current, the current waveform can be expressed by (7).

$$i' = \begin{cases} 0 & \psi \geq \psi_s \\ A'\cos(2\pi f_1 t + \alpha') + D'\exp(-t/T') & \psi < \psi_s \end{cases} \quad (7)$$

Three scenarios are discussed with (7).
1) Only inrush current
If the transformer is saturated, $i'$ is set to zero. And the amplitude of the non-inrush part is extremely low, which can be expressed in the form of fundamental and attenuation components ($A' \approx D' \approx 0$).

2) Serious internal fault current
If a serious internal fault occurs, the amplitude of fault current is much larger than that of inrush current. The effect of saturation is almost negligible, and as a result, the stage of $\psi \geq \psi_s$ would never exist.

3) Slight fault with inrush
When the number of short-circuit turns or grounding turns of transformer windings is small, the fault waveform and inrush current waveform will coexist in the original current. Then only the waveform of the non-inrush part, i.e., the slight fault current is retained. $A'$, $D'$ are the amplitudes of the fundamental and dc components of this weak fault.

This formula excludes the interference of other harmonic components in inrush current and retains the fundamental component. There are four unknown parameters in (7), which are $A'$, $\alpha'$, $D'$ and $T'$. In order to predict the values of the parameters in a known model, the NLS algorithm can be applied to solve the following equations:

$$\begin{cases} F(x) = \begin{bmatrix} i'[k_1] \\ \vdots \\ i'[k_n] \end{bmatrix} - \begin{bmatrix} A'\cos(2\pi f_1 t[k_1]+\alpha') + D'\exp(-t[k_1]/T') \\ \vdots \\ A'\cos(2\pi f_1 t[k_n]+\alpha') + D'\exp(-t[k_n]/T') \end{bmatrix}, \\ x = \arg\min_x \frac{1}{2}\|F(x)\|^2 \end{cases}$$

(8)

where $x = [A', \alpha', D', T']$, and only the non-zero sampling value of $i'$ and the corresponding sampling time are inputted.

The iterative methods, like the Newton-Raphson method, Gaussian iteration, and Levenberg-Marquardt (LM) algorithm, can solve such an NLS problem [27]. In this paper, the LM algorithm is adopted because it can avoid falling into local optimal solution, while the calculation speed is taken into account. The specific steps of the LM algorithm can be referred to [21]. After calculation, the extracted fundamental component for differential protection can be expressed as the phasor value: $\dot{I} = (A'/\sqrt{2})\angle\alpha'$.

### B. Overall Process of the proposed Protection Scheme

The overall process of the proposed method is shown in Fig. 5. This method mainly includes three modules, namely the non-inrush identification module, fundamental component extraction module, and ratio differential protection module. The specific operations of the proposed scheme are:

1) Firstly, the current waveforms on both sides of the transformer are collected. After normalization, the current signals are input into the inrush identification module. With the aid of data-driven technology, this module sets the waveform part disturbed by inrush current to 0, and only outputs the undisturbed waveform part. The output current waveform can be unitedly expressed by (7).

2) Next, the output current is imported into the fundamental component extraction module. According to the residual waveform, the NLS algorithm in this module can accurately estimate the four parameters in (7).

3) Finally, the estimated parameters of fundamental components are used to compute the differential current. The two phasor values on the transformer's both sides are imported into the ratio differential protection module to judge whether the relay trips or not. Its operation condition and the relevant calculation details are as follows:

$$\begin{cases} I_{op} \geq I_{op.0} \ \& \ I_{op} \geq KI_{res} \\ I_{op} = \left| \dot{I}_1 + \dot{I}_2 \right| \\ I_{res} = \left| \dot{I}_1 - \dot{I}_2 \right| \end{cases} \quad (9)$$

where $\dot{I}_1$, $\dot{I}_2$ are the phasor values of the fundamental component extracted by the proposed method, $I_{op.0}$ is the minimum operating current, $K$ is the restraint coefficient.

Compared with the traditional transformer differential protection, the proposed method doesn't need to configure inrush current blocking protection. It eliminates the negative impact of inrush current through the inrush identification module, which can avoid the protection delay caused by false locking. On the other hand, instead of the traditional DFT algorithm, the module of fundamental component extraction is applied to handle the incomplete waveform processed by the inrush identification module.

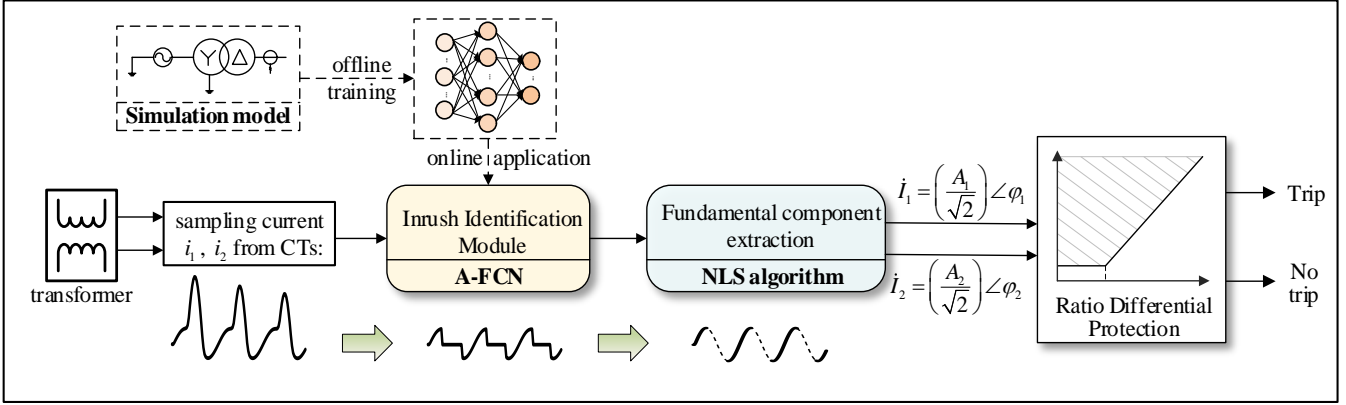

Fig. 5. Overall process of the proposed protection scheme

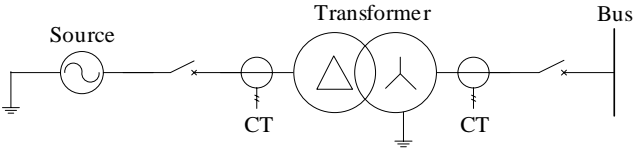

Fig. 6. Classical transformer model.

TABLE II
SPECIFIC PARAMETERS AND VARIABLE RANGES

| Parameter | |
|---|---|
| Ratio:13.8/242    Connection: △/ Y | |
| Closing winding: △    Fault winding: Y | |
| Capacity:160MVA    Iron loss: 0.01p.u.    Copper loss: 0.02p.u. | |
| **Variable** | |
| **Energizing state** | closing angle $\alpha$ : 0°, 30°, …, 330° |
| | remanence magnetism $\psi_r$ : 0%, 10%, …, 80% |
| | over-excitation multiples: 1.0, 1.1, 1.2, 1.3, 1.4, 1.5 |
| **Internal fault** | closing angle $\alpha$ : 0°, 30°,…, 330° |
| | short-circuit inter-turns: 0.8%, 1%, 2%, 4% |
| | location of fault windings: 10%, 30%,50%,70%,90% |
| | grounding turns: 3%, 5%, 7%, 10% |

\* The over-excitation multiple varies by adjusting the applied voltage.
\*\* The variables of the faulty transformer in the excitation state are composed of the above two conditions.

## IV. SIMULATION AND EXPERIMENTAL VERIFICATION

This section verifies the performance of A-FCN and the overall protection scheme by various simulation samples and the experimental data.

### A. Training of A-FCN

To construct a complete database for the data-aided method, a classical transformer model is built in the simulation software PSCAD/EMTDC, as shown in Fig. 6. As the figure shows, the close operation of the transformer is set on source-side, i.e. the delta-side. This kind of classical model simulates the nonlinearity of the iron core by injecting a compensation current. The sampling rate of this simulation system is 2000 Hz. Based on the generation mechanism of inrush current, the model parameters and variable ranges are given in Table II.

According to Table II, the simulation software conducts traversal scanning and generates 8600 samples in total. To improve the adaptability of the network, three main types mentioned in Section III are randomly divided in a ratio of 7:3. The final training set contains 6020 samples and the test set has 2580 samples. The training iteration epoch is 300. During the training process, the early stopping method is used to prevent overfitting. If the loss function doesn't further decrease in the next 30 epochs, the training is stopped and the network model corresponding to the local minimum point is saved. Here the loss function is mean square error (MSE) defined as (10).

$$\text{loss} = \sum_{k=1}^{n} (S[k] - S'[k])^2 / n \quad (10)$$

where $S'$ is the predicted label of network output. Considering the recognition speed, the time windows is 20*ms*.

The loss curves of A-FCN are shown in Fig. 7. The two curves converge quickly and tend to be stable after about 60 epochs. According to early stopping method, the loss value of the testing set reaches the lowest point in the 146[th] epoch, when the testing performance of the network is best and there is no overfitting.

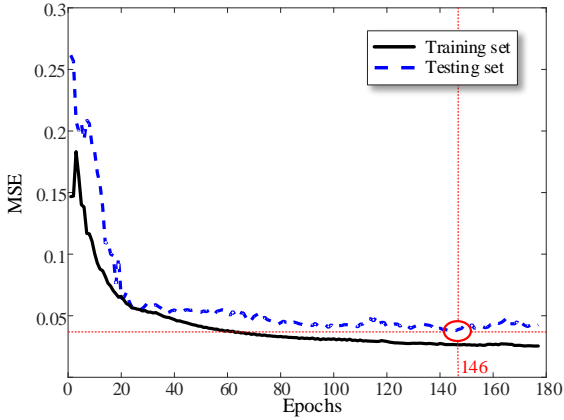

Fig. 7. Loss curves of network training and testing.

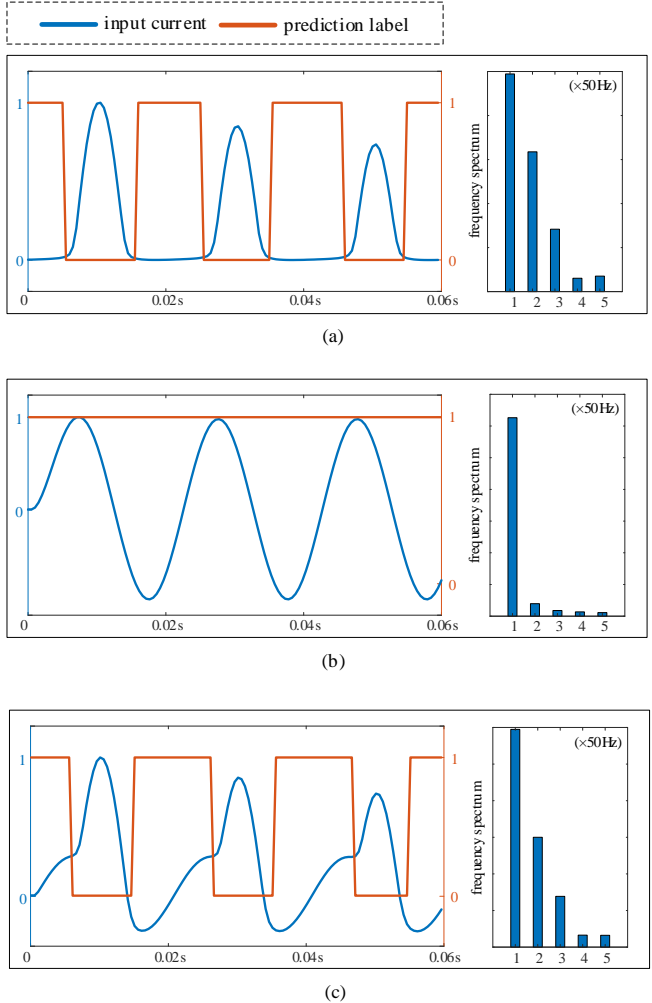

Fig. 8. Identification results and frequency spectrums of the three scenarios: (a) inrush current, (b) internal fault current, (c) slight fault with inrush.

B. *Accuracy and generalization verification of inrush identification module*

The main purpose of the A-FCN is to distinguish the scenarios of inrush current, internal fault, and slight fault with inrush. The identification results of the three scenarios are shown in Fig. 8. It can be seen that the output label is 0 when the corresponding waveform part is distorted by inrush current, while the undistorted part of the waveform is marked as 1 in three scenarios. Noted that when a serious fault occurs as shown in Fig. 8 (b), the fault waveform masks the inrush, so the labels are always one. Besides, for the concerned scenario of weak fault current with inrush, A-FCN can accurately distinguish the inrush and non-inrush parts. From the perspective of the frequency domain, only the fundamental component exists in the case of severe internal fault, as shown in Fig. 8(b). However, the frequency spectrum of the slight fault with inrush is similar to that of inrush current, whose second harmonic content is very high as shown in Fig.8(a) and (c). The similarity of these two scenarios easily leads to the protection false blocking. The waveform marked as 1 can be considered to only contain the fundamental frequency component, so extracting it by A-FCN can constitute a protection strategy that isn't interfered by inrush current.

In order to verify the generalization performance of the proposed A-FCN, more simulation results and experimental data are provided. The test data includes:

1) *Simulation data with different transformer model*: The test transformer is the UMEC model. Its copper loss and iron loss are 0.05 and 0.02 p.u. respectively. Unlike the classical model, the UMEC model simulates core saturation by U-I curve. In this paper, two U-I curves are set to simulate different characteristics of the transformer, whose parameters are given in Table III. The measuring CTs adopt the JA model in PSCAD. The simulation testing set includes 1012 inrush current samples, 917 internal faults, and 862 samples of slight faults with inrush.

TABLE III
SPECIFIC PARAMETERS OF U-I CURVES

| Curve 1 | | Curve 2 | |
|---|---|---|---|
| Voltage(p.u.) | Current(%) | Voltage(p.u.) | Current(%) |
| 0.00 | 0.00 | 0.00 | 0.00 |
| 0.40 | 0.18 | 0.10 | 0.005 |
| 0.60 | 0.49 | 0.50 | 0.026 |
| 1.00 | 0.98 | 1.00 | 0.0605 |
| 1.25 | 2.00 | 1.10 | 0.42 |
| 1.28 | 3.10 | 1.40 | 36.29 |
| 1.36 | 6.52 | 1.60 | 78.09 |
| 1.45 | 20.54 | 1.80 | 125.83 |
| 1.50 | 60.22 | 1.90 | 151.15 |
| 2.50 | 124.39 | 2.00 | 177.12 |

2) *Experimental data*: The experimental 6kVA three-phase transformer group is composed of three single-phase transformers with two windings, and the connection is Y/Δ-11. Due to the limitation of the experiment condition, the rated voltage is 380V/220V. The measurement CTs adopt Hall sensors, and the data measured by CTs is transferred to the computer via a data recorder. The experimental schematic and layout diagrams are present in Fig. 9. In the no-load closing test, 223 sets of experimental data are obtained by setting different fault locations and short-circuit turns.

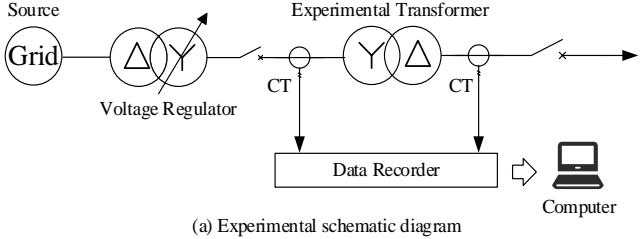

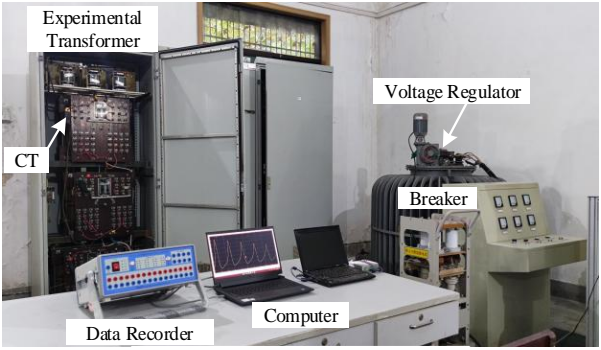

Fig. 9. Experimental setup of the transformer.

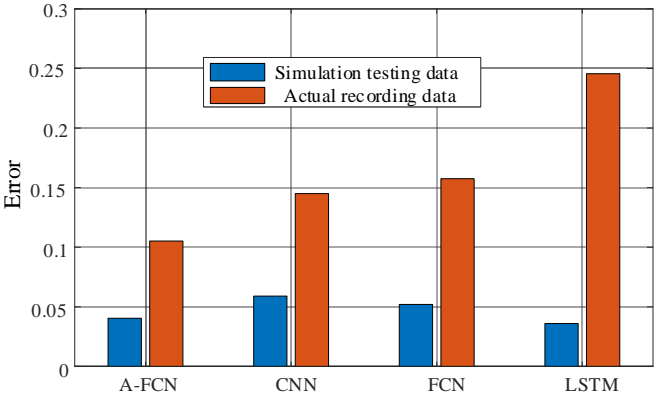

Fig. 10. Loss index of performance evaluation.

Based on the testing samples, the comparison is made between A-FCN and other data-driven methods including ordinary CNN, FCN, and long-short-term memory network (LSTM). The results of these networks are shown in Fig. 10, where the error is the loss function defined in (10).

For the simulation test set, all four neural networks perform well, of which LSTM slightly prevails and A-FCN is second. But for the experimental data, the error of A-FCN is lower than that in other networks. It's worth mentioning that LSTM has the worst results on the experimental data despite its good performance in simulation tests. In conclusion, the proposed A-FCN is better than other methods in dealing with more complex recording waveforms.

Using the loss of the testing datasets as an evaluation index reflects the prediction accuracy, that is, the ratio of the correctly classified sampling points to the total sampling points. However, if the sample types are unbalanced, the high accuracy is likely to be due to the preference for a certain type, which cannot reasonably reflect the prediction ability of the network model for all categories. For the datasets in this test, the number of sampling points in the non-inrush part is not balanced with that in inrush part. Therefore, this paper further uses precision, recall, and f1-score to comprehensively evaluate the performance, as shown in Table IV. The sampling points of non-inrush parts are treated as positive samples, and inrush parts are regarded as negative samples. Precision is the proportion of the actual positive samples in the predicted positive samples. Recall is the proportion of the predicted positive samples in the actual positive samples. F1-score is the harmonic mean of Precision and Recall.

TABLE IV
CLASSIFICATION INDEX OF PERFORMANCE EVALUATION

|  | A-FCN | CNN | FCN | LSTM |
|---|---|---|---|---|
| Accuracy | 91.78% | 88.46% | 89.46% | 83.05% |
| Precision | 90.12% | 86.02% | 88.01% | 83.71% |
| Recall | 96.92% | 84.82% | 94.00% | 85.58% |
| F1 score | 93.20% | 89.96% | 90.45% | 80.11% |

The above table is the result of the common statistics of the simulation data and experimental data. Since the number of the experimental data is less than the simulation data, the discriminant results are scaled proportionally according to the ratio of the two datasets. It can be seen from Table III that the accuracy and generalization ability of A-FCN are higher than those of CNN and FCN after the attention mechanism is introduced, and each index exceeds 90%. Therefore, it's concluded that A-FCN proposed in this paper has better accuracy and generalization performance.

*C. Effectiveness Verification of the fundamental component extraction module*

This part verifies the effectiveness of the fundamental component extraction module. Figure 11-14 show the results of the module under different simulation scenarios. Moreover, the proposed module is compared with the traditional DFT algorithm. In each figure, the red line in (a) is the fundamental component extracted by the data-aided module while the dark line represents the actual waveform. In (b), $I_{dft}$ and $I_{ext}$ are the effective values of the fundamental component calculated by the DFT algorithm and the proposed method respectively, and $I_{tru}$ is the true fundamental component in transformer current.

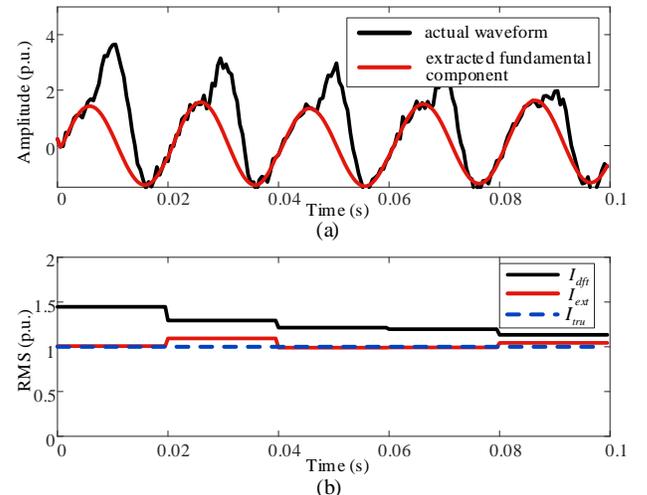

Fig. 11. (Case 1) Slight fault with inrush current: (a) actual & extracted waveforms, (b) actual & estimated RMS.

In Fig. 11, case 1 is the slight internal fault current with inrush. It can be seen that, in the first period, $I_{dft}$ is roughly 1.5 times of the real fundamental component because of inrush interference, and it gradually approaches $I_{tru}$ until the inrush current decays for 5 periods. But the proposed module can accurately estimate the fundamental component as the two lines representing $I_{ext}$ and $I_{tru}$ basically coincide. Besides, the noise contained in the current waveform doesn't affect the extraction effect, so the proposed method has a certain anti-noise ability.

Symmetrical inrush, as shown in Fig. 12, doesn't have the unilateral characteristics of the typical inrush current, and its similarity with fault current is higher, so it's more difficult to distinguish. As a result, $I_{dft}$ is much higher than the actual fundamental component. By contrast, $I_{ext}$ is calculated from the non-inrush part, so it's close to $I_{tru}$.

When the amplitude of inrush current is too high, it may lead to the saturation of the measuring CT, and the waveform suffers secondary distortion, as shown in Fig. 13. The waveform's peak is weakened and the reverse inrush occurs, making the waveform more complicated. As a result, $I_{dft}$ varies considerably from the true value, especially in the first period. By contrast, the proposed method can still accurately extract the real fundamental component.

Over-excitation is another normal saturation state, which is caused by the increase in magnetic flux with the voltage or frequency changing [28]. The second harmonic content of the over-excitation inrush current is usually less than that of the ordinary excitation inrush current, while the third harmonic content increases, so the scheme relying on the second harmonic braking may malfunction. As shown in the Fig. 14, this over-excitation waveform is the line current of the delta winding. The waveform in the saturation region becomes more complex, while the effective information in the unsaturation region decreases. It can be seen from the extraction effect of the fundamental wave that the proposed method still performs well in the scenario of transformer core overexcitation.

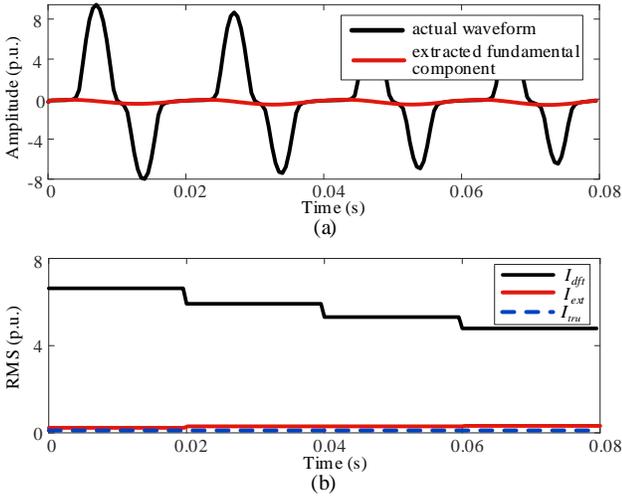

Fig. 12. (Case 2) Symmetrical inrush current: (a) actual & extracted waveforms, (b) actual & estimated RMS.

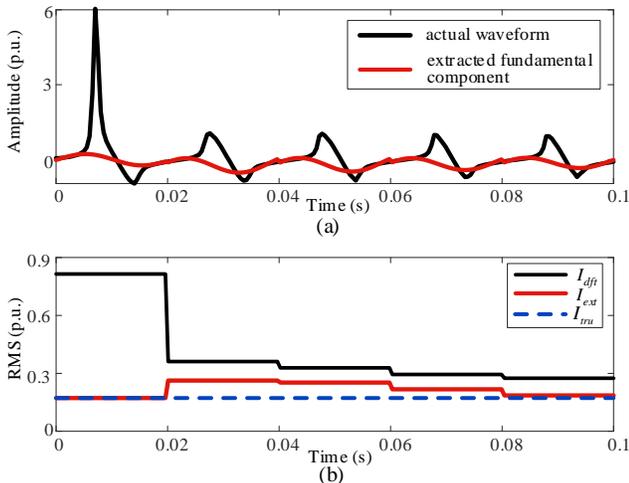

Fig. 13. (Case 3) Inrush current with CT saturation: (a) actual & extracted waveforms, (b) actual & estimated RMS.

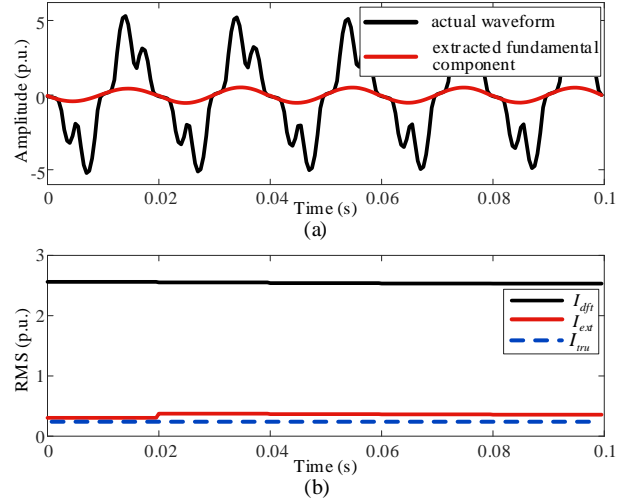

Fig. 14. (Case 4) Inrush current with over-excitation: (a) actual & extracted waveforms, (b) actual & estimated RMS.

To conclude, the proposed fundamental component extraction module has better performance than the conventional DFT algorithm, especially in complex transformer's current waveforms. This superiority can be attributed to the combination of this module and the inrush identification module. The inrush identification module can distinguish and retain the non-inrush part, which significantly reduces the interference of inrush currents on fundamental component extraction.

*D. Performance of the proposed protection scheme*

To prove the superiority of the proposed method in the real-time application, this paper uses the experimental data to compare it with the conventional ratio differential protection. After standardization, $K$ is 0.7, $I_{op.0}$ is 0.3 p.u., and the minimum restraint current is 1 p.u. in (9). The compared differential protection is equipped with the DFT algorithm and an SHR module whose threshold is set to 0.2. The data windows of both methods are one cycle (20ms), and the data window contains 40 sampling points at the sampling rate of 2kHz. The average calculation time of the DFT algorithm in the traditional method is 1.03ms, and that of the proposed method is 8.70ms through testing. Therefore, the updating time of the sliding window of the above two methods is set to 2ms and 10ms respectively. The above calculation is performed on the computer with 16GB RAM and i7-10700 CPU.

As shown in Fig. 15(a), the power transformer with 4.54% inter-turns fault closes without load, and the effective value of the fault component is about 2.37 p.u., which is covered by the inrush current with a higher amplitude. In Fig. 16(b), SHR remains above 0.2 in the first 11 cycles, and the conventional protection is blocked, which causes the fault to last 0.22s. But in Fig. 15(a), the fundamental component extracted by the proposed method is basically coincident with the original waveform without inrush interference, and this fundamental component reflects the actual fault current. Therefore, the protection can operate correctly within 30ms by the proposed method.

A slighter fault case is further discussed in Fig. 16(a). Only 2.27% inter-turns fault occurs in the transformer, even after 14 periods' attenuation, SHR is still larger than 0.3 and the differential protection keeps blocked. By contrast, the proposed scheme can quickly extract the real fundamental component even in the first period. In Fig. 16(a), the effective value of the extracted fundamental component is 0.73 p.u., which is greater than the minimum operating current, then the protection can operate in time.

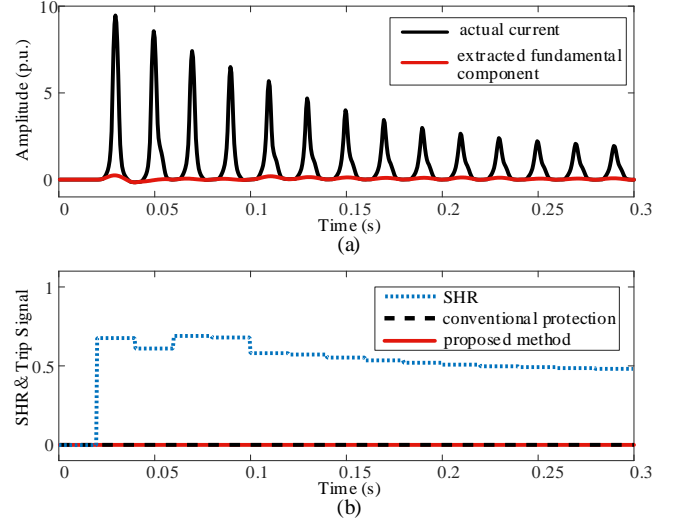

Fig. 17. (Case 7) Inrush current without fault: (a) waveforms of differential current, (b) SHR & trip signals

Figure 17(a) shows the typical waveform of inrush current. If the transformer core is saturated, the current peak occurs on one side. On the contrary, the amplitude of the non-inrush part is very small when the transformer is under unsaturation. In such a case, the traditional protection and the proposed protection can both avoid false tripping.

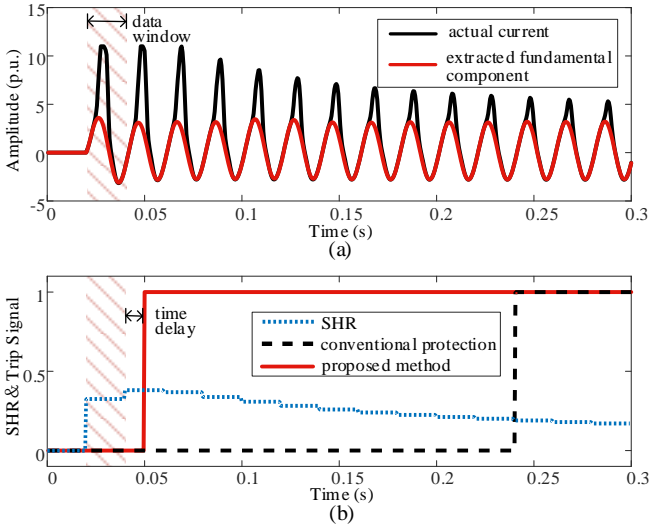

Fig. 15. (Case 5) 4.54% inter-turns fault with inrush current: (a) waveforms of differential current, (b) SHR & trip signals.

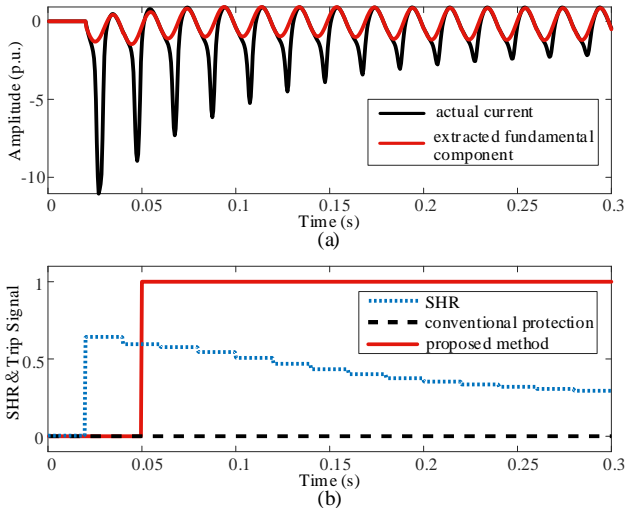

Fig. 16. (Case 6) 2.27% inter-turns fault with inrush current: (a) waveforms of differential current, (b) SHR & trip signals.

TABLE IV
TRIPPING TIME COMPARISON OF DIFFERENT PROTECTION

|  | Details | Conventional protection | Proposed method |
|---|---|---|---|
| Case1 | Slight fault with inrush | 0.14s | 0.03s |
| Case2 | Symmetrical inrush current | No trip | No trip |
| Case3 | Inrush current with CT saturation | No trip | No trip |
| Case4 | Inrush current with over-excitation | False trip | No trip |
| Case5 | Slight fault with inrush | 0.22s | 0.03s |
| Case6 | Slight fault with inrush | >0.3s | 0.03s |
| Case7 | Typical inrush current | No trip | No trip |

Table IV summarizes the tripping time in different cases. The conventional protection and the proposed scheme will not trip in no-fault scenarios, which include the symmetrical/

asymmetrical inrush current, over-excitation current and the inrush current with CT saturation in cases 2, 3, 4 and 7. However, for a slight fault current distorted by inrush in cases 1, 5, and 6, the conventional protection will be blocked due to the SHR criterion, which may delay the fault removal. Such delays may cause the slight faults to develop into serious ones, causing damage to the transformers as well as the system. By contrast, the proposed method can accurately detect the slight fault hidden in inrush current and trip within 0.03s in cases 1, 5, and 6, which greatly improves the sensitivity and rapidity of differential protection.

## V. Conclusion

In this paper, a data-aided power transformer differential protection without inrush blocking module is proposed. The proposed protection consists of the data-driven inrush identification module, NLS-based fundamental component extraction module, and the ratio differential current computation module. The combination of the inrush identification module and the fundamental component extraction module can accurately estimate the differential current from the non-inrush part of the current waveform, so the proposed method is immune to inrush interference and doesn't need an inrush blocking module. The simulation and experimental results show that A-FCN can effectively distinguish between the inrush and non-inrush parts, which has stronger generalization than other neural networks. What's more, for complex simulation and experimental data, the proposed method can extract the fundamental component from the slight fault current hidden in inrush, which realizes the fast operation of differential protection. Compared with the conventional blocking strategy, the proposed method significantly improves the sensitivity and rapidity of transformer differential protection to slight faults.

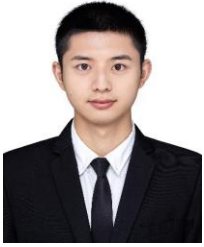

**Zexuan Lin** received the B.S. degree from Shandong University, Jinan, China, in 2016. He is currently working toward M.S. degree in Xi'an Jiaotong University, Xi'an, China. His research intersests is the application of artificial intelligence in the relay protection.

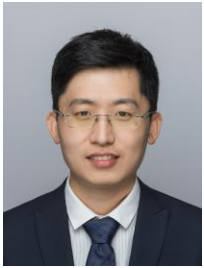

**Songhao Yang** (S'18-M'19) was born in Shandong, China, in 1989. He received the B.S. and Ph.D. degrees in electrical engineering from the Xi'an Jiaotong University, Xi'an, China, in 2012 and 2019, respectively. Besides, he received the Ph.D. degree in electrical and electronic engineering from Tokushima University, Japan, in 2019.

Currently, he is an Assistant Professor at Xi'an Jiaotong University. His research interest includes power system control and protection.

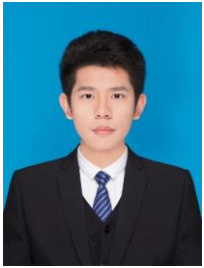

**Yubo Zhang** (S'21) received the B.S. degree in electrical engineering from Xi'an Jiaotong University, China, in 2019. He is currently pursuing the Ph.D. degree in the school of Electrical Engineering in Xi'an Jiaotong University. His main field of interest includes the power system control and protection.

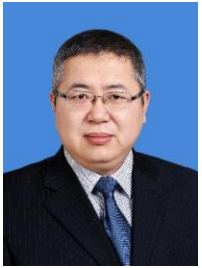

**Zhiguo Hao** (M'10) was born in Ordos, China, in 1976. He received his B.Sc. and Ph.D. degrees in electrical engineering from Xi'an Jiaotong University, Xi'an, China, in 1998 and 2007, respectively. He has been a Professor with the Electrical Engineering Department, Xi'an Jiaotong University. His research interest includes power system protection and control.

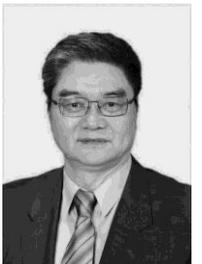

**Baohui Zhang** (SM'99-'F'19) was born in Hebei Province, China, in 1953. He received the M.Eng. and Ph.D. degrees in electrical engineering from Xi'an Jiaotong University, Xi'an, China, in 1982 and 1988, respectively. He has been a Professor in the Electrical Engineering Department at Xi'an Jiaotong University since 1992. His research interests are system analysis, control, communication, and protection.